# Structural features and tetragonal-orthorhombic phase transition in $SrFe_2As_2$ and $CaFe_2As_2$

C. Ma, H.X. Yang, H.F. Tian, H.L Shi, J.B. Lu, Z.W. Wang, L.J. Zeng, G.F. Chen, N.L. Wang, and J.Q. Li*,

*Beijing National Laboratory for Condensed Matter Physics, Institute of Physics,*

*Chinese Academy of Sciences, Beijing 100080, China*

*Corresponding author. E-mail: <u>LJQ@aphy.iphy.ac.cn</u>

The structural properties of the $SrFe_2As_2$ and $CaFe_2As_2$ compounds have been extensively analyzed by transmission electron microscopy (TEM) from room temperature down to 20K. The experimental results demonstrate that the $SrFe_2As_2$ crystal, in consistence with previous x-ray data, has a tetragonal structure at room temperature and undergoes a tetragonal (T)-orthorhombic (O) phase transition at about 210K. Moreover, twinning lamella arising from T-O transition evidently appears in the orthorhombic phase. On the other hand, TEM observations of $CaFe_2As_2$ reveal the presence of a pseudo-periodic structural modulation with the periodicity of around 40nm at room temperature. This modulation is likely in connection with the local structural distortions within the Ca layer. In-situ cooling TEM observations of $CaFe_2As_2$ reveal the presence of complex domain structures in the low-temperature orthorhombic phase.

Following the discovery of superconductivity at Tc ≈ 26 K in LaFeAsO(F) [1], a variety of ZrCuSiAs-type (1111 phase) superconductors with Tc ranging from 20K to 55K have been discovered, such as, $RFeAsO_{1-x}F_x$ [2-5], $RFeAsO_{1-\delta}$ [6] and $R_{1-x}Th_xFeAsO$ [7,8] (R = La, Ce, Pr, Nd, Sm, Gd, Tb; Th= Sr, Ca). Similar to the cuprate high-temperature superconductors, the $(Fe_2As_2)$-layers are considered to be the conducting layers playing a critical role on the occurrence of superconductivity, while the $(R_2O_2)$-layers inject charge carriers to the former via chemical doping and also retain structural integrity of the $(Fe_2As_2)$-layers. Furthermore, the $ThFe_2As_2$ (Th = Ca, Sr, Ba, Eu) materials with the $ThCr_2Si_2$-type structure (122 phase) are found to be superconductors under certain conditions. Superconductivity in this family can be induced by hole doping up to 38 K, e.g. partial substitution of K or Na for Ca, Ba, or Sr atoms, and can be also induced by applied hydrostatic pressure [9-14].

Recent structural and physical investigations have documented that the (FeAs)-based system contains noteworthy competition among numerous ordered states along with doping charge carriers. The ROFeAs and $ThFe_2As_2$ parent samples in general show remarkable anomalies between 120 K and 220 K as clearly observed in the measurements of electrical resistivity and magnetic susceptibility, as shown in Fig. 1 [15,16]. These anomalous properties are found to be in connection with a spin density wave (SDW) instability and a T-O phase transition at low temperature [17-20]. Measurements of structural and physical properties on $BaFe_2As_2$, $SrFe_2As_2$, and $CaFe_2A_s$ single crystalline samples suggested the phase transitions in the materials occur respectively at about 140K, 205 K and 170K. Neutron and x-ray diffraction studies demonstrated that this low-temperature phase transition can be well characterized as tetragonal I4/mmm (T) to orthorhombic Fmmm (O) structure [21-28]. In order to directly reveal the microstructure futures and structural changes in association with this phase transition, we have carried out extensive in-situ TEM observations on $ThFe_2As_2$ samples from room temperature down to 20K. It was found that the $CaFe_2As_2$ crystal in general shows more complex microstructure properties in comparison with $SrFe_2As_2$. In this paper, we will report our TEM results obtained in recent investigations, a pseudo-periodic modulation $CaFe_2As_2$ and structural twinning arising from T-O phase transition in $ThFe_2As_2$ crystals will be systematically analyzed.

The $ThFe_2As_2$ (Th=Sr, Ca, and Ba) single crystals used in present study were synthesized by methods as reported in previous publications [16,29]. Specimens for TEM observations were prepared by peeling off a very thin sheet of a thickness around several tens microns from the single crystal and then followed by ion milling. In addition, we also prepared some thin samples for electron diffraction experiments simply by peeling off fine fragments from the layered single crystal which were dissolved in chloroform and then supported by a copper grid coated with a thin carbon film. Microstructure analyses were performed on a FEI Tecnai-F20 TEM equipped with double-tilt cooling holders.

We firstly focus on the microstructure and changes of structural properties in association with the low-temperature T-O phase transition ($T_{TO}$~205K) in the $SrFe_2As_2$ crystal. Previous structural characterizations by means of both x-ray diffraction and TEM observations revealed that the $SrFe_2As_2$ sample has a tetragonal structure with a space group I4/mmm at room temperature. Fig. 2 shows a schematic structural mode, illustrating the layered structural feature of $ThFe_2As_2$. Fig. 3 displays a high-resolution TEM image taken from a $SrFe_2As_2$ sample along the [100] zone-axis direction. This image was obtained from a thin area nearly under the Scherzer defocus, the dark dots, therefore, represent the atomic positions as clearly indicated in the structural model superimposed on the experimental image. High-resolution TEM image simulations for $SrFe_2As_2$ were carried out by varying the crystal thickness from 2 to 5 nm and the defocus value from -30 to -60 nm. A simulated image with the defocus value of -45nm and the thickness of 3nm is superimposed on the experimental image and appears to be in good agreement with the experimental one.

On the other hand, it is noted that the distance between Fe and As column (~0.14nm) in present case is noticeably smaller than the spatial resolution of our TEM (0.2nm). Hence, the positions of Fe and As atoms can be not clearly distinguished in this experimental image. A further structural study using a Cs-corrector TEM is under progress for directly characterizing the atomic structural feature within a FeAs layer.

The $SrFe_2As_2$ compound in general shows a structural phase transition at about 205 K associated with spin density wave instability [16,20]. In order to directly observe the structural changes following this phase transition, we have performed a series of in-situ TEM investigations on well-characterized single crystalline $SrFe_2As_2$ samples. TEM observations revealed that the electron diffraction pattern obtained in the low-temperature orthorhombic phase shows up unambiguous spot splitting due to the structural transformation. Moreover, notable structural changes can be also recognized in both crystal symmetry and microstructure. Fig. 4(a)-(d) show the bright field TEM images and corresponding electron diffraction patterns taken along the [001] zone-axis direction at 300K and 100K, respectively, illustrating the typical structural changes in $SrFe_2As_2$ through the T-O phase transition. Similar structural change was also observed in the $CaFe_2As_2$ sample as discussed in the following context. Moreover, microstructure observations demonstrated that the low-temperature orthorhombic $SrFe_2As_2$ is heavily twinned with the $(110)_{orth}$ plane being its twining plane. The a- and b-orthorhombic axes are mirrored across the twin boundaries. Fig. 5(a) shows a typical bright-field TEM image illustrating the twining lamella in a $SrFe_2As_2$ crystal. The thickness of twinning domains, often depending visibly on temperature, ranges from 100 to 400nm at 100K. Actually there are many factors that could affect the twin density, such as thermal treatments, impurity content, sample preparation techniques. The exact atomic structure of the twin boundaries in $SrFe_2As_2$ has not yet been determined. Considering the simultaneous occurrence of the SDW instability of Fe ions and T-O phase transition, we propose a schematic structural model illustrating twins within a Fe layer in Fig. 5(b). The [001] zone-axis electron diffraction pattern obtained from several twin domains can clearly show the difference in the *a* and *b* lattice parameters as spot splitting along the $<110>_{Orth}$ direction, as shown in Fig. 5(c) and schematically illustrated in Fig. 5(d). The orthorhombic splitting angle in present case, estimated from the electron diffraction, is given by $\alpha = 2(tg^{-1}(a/b) - 45°) \approx 1°$, where a and b are lattice parameters for the orthorhombic phase. Therefore, the ratio of the lattice parameter a/b is expected to be about 1.009 which is close to the value from the x-ray and neutron scattering ($\approx$1.010) [20].

We now proceed to study the microstructure and structural phase transition in $CaFe_2As_2$. Fig. 6(a) shows a bright-field TEM image taken from a $CaFe_2As_2$ crystal at room temperature. The striking structural feature, in contrast with that of $SrFe_2As_2$, is the appearance of a pseudo-periodic modulation along the [110] direction, the wave length of this modulation is estimated to be about 40nm. The periodicity of this modulation usually does not vary rapidly from one area to another. Regular contrast of modulation stripes can be clearly observed in many crystals. Moreover, this kind of modulation stripes can commonly coexist along two equivalent [110] and [-110] directions, and two orthogonal modulations overlap (meet) to yield an interesting pattern as shown in Fig. 6(b). This type of microstructure feature of lenticular domains can be called as "tweed" structure as commonly observed in many alloys [30] and Fe-doped $YBa_2Cu_3O_{7-x}$ superconductor [31,32], in which the tweed structure arises from the compositional disorder or local lattice distortions. Insets show the data of the fast Fourier transformation (FFT) for the TEM images, in which only the superstructure reflections following the direct spot are shown. Our careful structural analysis demonstrate that this pseudo-periodic structural modulation only appears in $CaFe_2As_2$, neither $BaFe_2As_2$ nor $SrFe_2As_2$, considering the relatively smaller size of $Ca^{2+}$ (1.0Å) ion in comparison with $Sr^{2+}$(1.26Å) and $Ba^{2+}$ (1.46Å) ions. We suggest that this modulation arises essentially from local structural fluctuations within the Ca atomic layers, which results in a visible stripe pattern going along the diagonal of Ca squares. It is remarkable that the Bragg spot at the systematic (100) position, which should be absent according to the space group I4/mmm, is obviously observed in the electron diffraction pattern (Fig. 7(a)). The presence of the (100) reflection spot can not be attributed to the multiple diffraction effects, but being in correlation with a notable alteration of local crystal symmetry in association with the structural modulation and the tweed structural domains as well. Because no evident spots splitting or streaking in the electron diffraction is observed at room temperature, the local distortion in $CaFe_2As_2$ should be rather small within the a-b plane.

Like the $SrFe_2As_2$ sample, the $CaFe_2As_2$ sample also adopts an orthorhombic structure at low temperature, with the T-O phase transition occurring at the temperature of about 170K. Our in-situ cooling TEM study demonstrated that this T-O transition is also well characterized by the unambiguous spot splitting (Fig. 7(b)) and the formation of twin boundaries (Fig. 8(a)). The sharp contrast of parallel lines along the $[110]_{orth}$ direction are twin boundaries. Actually, the low-temperature TEM image often exhibits complex microstructures with coexistence of twin domain structure and a tweed structure. The twin boundary makes an angle of about 45º with the pseudo-periodic modulation, as shown in the schematic illustration (Fig. 8(b)). The twin boundary in orthorhombic phase goes along the diagonal of the iron squares, i.e. the $[110]_{orth}$ or $[-110]_{orth}$ direction. The width of the domain (100~500nm) is much larger than that of the modulation stripes (~40nm). These structure features are much different from what are observed in $YBa_2Cu_3O_{7-x}$, in which the strain fluctuations show well developed embryos of the tweed texture and the tweed transforms into multitwin structure in the $CuO_2$ layers below the T-O transition temperature [33]. However, our TEM observations of $CaFe_2As_2$ showed that no visible changes of tweed structure happen through T-O transformation at 170K. This fact suggests that the twin lamella and tweed structure coexisting at low temperature occurs essentially in different atomic layers, i.e. Fe layer for twin and Ca layer for tweed, respectively.

Although the $CaFe_2As_2$ crystal lattice is locally distorted within the Ca layers recognized as one-dimensional stripes or the tweed pattern, the average crystal lattice as determined by electron diffraction can also be considered as a tetragonal structure, which is in agreement with the x-ray and neutron diffraction experimental data at room temperature. Taking into account of the considerably different physical properties observed in the $CaFe_2As_2$ and $SrFe_2As_2$ materials, it is likely that the pseudo-periodic structural modulation in $CaFe_2As_2$ could have certain impact on the structure instability and physical properties. For instance, the anomalies in $CaFe_2As_2$ associated with the SDW instability in both electrical resistivity and magnetic susceptibility show visibly dissimilarities with what observed in $SrFe_2As_2$ (Fig. 1).

In summary, we have extensively investigated the structural properties and phase transitions in $ThFe_2As_2$ (Th=Ca, Sr) compounds. The $SrFe_2As_2$ crystal, in consistence with previous x-ray and neutron diffraction data, has a tetragonal structure at room temperature, and twin domains from T-O phase transition evidently appear in the orthorhombic phase. Moreover, TEM observations of $CaFe_2As_2$ reveal the presence of a new pseudo-periodic structural modulation at room temperature, and this modulation with a wave length of around 40nm can be briefly interpreted as the local structural fluctuation within the Ca layers. In the low-temperature orthorhombic phase, the coexistence of twin lamella and tweed structure yields complex microstructure features in the a-b plane of the $CaFe_2As_2$ crystals.

Acknowledgments

This work is supported by the National Science Foundation of China, the Knowledge Innovation Project of the Chinese Academy of Sciences, and the 973 projects of the Ministry of Science and Technology of China.

References

[1] Y. Kamihara, T. Watanabe, M. Hirano, and H. Hosono, J. Am. Chem. Soc. 130, 3296 (2008).

[2] X.H. Chen, T. Wu, G. Wu, R.H. Liu, H. Chen, and D.F. Fang, Nature 453, 761 (2008).

[3] G.F. Chen, Z. Li, G. Li, J. Zhou, D. Wu, J. Dong, W.Z. Hu, P. Zheng, Z.J. Chen, J.L. Luo, and N.L. Wang, Phys. Rev. Lett. 100, 247002 (2008).

[4] Z.A. Ren, J. Yang, W. Lu, W. Yi, G.C. Che, X.L. Dong, L.L. Sun, and Z.X. Zhao, Materials Research Innovations 12, 106 (2008).

[5] Z.A. Ren, J. Yang, W. Lu, W. Yi, X.L. Shen, Z.C. Li, G.C. Che, X.L. Dong, L.L. Sun, F. Zhou, and Z.X. Zhao, Europhys. Lett. 82, 57002 (2008).

[6] Z.A. Ren, G.C. Che, X.L. Dong, J. Yang, W. Lu, W. Yi, X.L. Shen, Z.C. Li, L.L. Sun, F. Zhou, and Z.X. Zhao, Europhys. Lett. 83, 17002 (2008).

[7] C. Wang, L. Li, S. Chi, Z. Zhu, Z. Ren, Y. Li, Y. Wang, X. Lin, Y. Luo, X. Xu, G. Cao, and Z.A. Xu, Europhys.Lett. 83, 67006 (2008).

[8] L.J. Li, Y.K. Li, Z. Ren, Y.K. Luo, X. Lin, M. He, Q. Tao, Z.W. Zhu, G.H. Cao, and Z.A. Xu, arXiv:0806.1675v2.

[9] M.S. Torikachvili, S.L. Bud'ko, N. Ni, and P.C. Canfield, Phys. Rev. Lett. 101, 057006 (2008).

[10] M. Rotter, M. Tegel, and D. Johrendt, Phys. Rev. Lett. 101, 107006 (2008).

[11] A.S. Sefat, R. Jin, M.A. McGuire, B.C. Sales, D.J. Singh, and D. Mandrus, Phys. Rev. Lett. 101, 117004 (2008).

[12] K. Sasmal, B. Lv, B. Lorenz, A.M. Guloy, F. Chen, Y.Y. Xue, and C.W. Chu, Phys. Rev. Lett. 101, 107007 (2008).

[13] H.S. Jeevan, Z. Hossain, D. Kasinathan, H. Rosner, C. Geibel, and P. Gegenwart, Phys. Rev. B 78, 092406 (2008).

[14] G. Li, W. Z. Hu, J. Dong, Z. Li, P. Zheng, G.F. Chen, J.L. Luo, and N.L. Wang, Phys. Rev. Lett. 101, 107004 (2008).

[15] N. Ni, S. Nandi, A. Kreyssig, A.I. Goldman, E.D. Mun, S.L. Bud'ko, and P.C.

Canfield, Phys. Rev. B 78, 014523 (2008).

[16] G.F. Chen, Z. Li, J. Dong, G. Li, W.Z. Hu, X.D. Zhang, X.H. Song, P. Zheng, N.L. Wang, and J.L. Luo, Chin. Phys. Lett. 25, 3403 (2008).

[17] J. Dong, H.J. Zhang, G. Xu, Z. Li, G. Li, W.Z. Hu, D. Wu, G.F. Chen, X. Dai, J.L. Luo, Z. Fang, and N.L. Wang, Europhys. Lett. 83, 27006 (2008).

[18] C. de la Cruz, Q. Huang, J.W. Lynn, Jiying Li, W. Ratcliff, J.L. Zarestky, H.A. Mook, G.F. Chen, J.L. Luo, N.L. Wang, and P. Dai, Nature 453, 899 (2008).

[19] T. Yildirim, Phys. Rev. Lett. 101, 057010 (2008).

[20] J. Zhao, W. Ratcliff, II, J.W. Lynn, G.F. Chen, J.L. Luo, N.L. Wang, J. Hu, and P. Dai, Phys. Rev. B 78, 140504(R) (2008).

[21] M. Tegel, M. Rotter, V. Weiß, F.M. Schappacher, R. Pöttgen, and D. Johrendt, J. Phys.: Condens. Matter 20, 452201 (2008).

[22] M. Rotter, M. Tegel, D. Johrendt, I. Schellenberg, W. Hermes and R. Pöttgen Phys. Rev. B 78, 020503(R) (2008).

[23] Q. Huang, Y. Qiu, Wei Bao, J.W. Lynn, M.A. Green, Y.C. Gasparovic, T. Wu, G. Wu, and X.H. Chen, arXiv:0806.2776v1.

[24] J.-Q. Yan, A. Kreyssig, S. Nandi, N. Ni, S.L. Bud'ko, A. Kracher, R.J. McQueeney, R.W. McCallum, T.A. Lograsso, A.I. Goldman, and P.C. Canfield, Phys. Rev. B 78, 024516 (2008).

[25] N. Ni, S. Nandi, A. Kreyssig, A.I. Goldman, E.D. Mun, S.L. Bud'ko, and P.C. Canfield, Phys. Rev. B 78, 014523 (2008).

[26] A.I. Goldman, D.N. Argyriou, B. Ouladdiaf, T. Chatterji, A. Kreyssig, S. Nandi, N. Ni, S.L. Bud'ko, P.C. Canfield, and R.J. McQueeney, Phys. Rev. B 78, 100506(R) (2008).

[27] R. Mittal, Y. Su, S. Rols, T. Chatterji, S.L. Chaplot, H. Schober, M. Rotter, D. Johrendt, and Th. Brueckel, Phys. Rev. B 78, 104514 (2008).

[28] A.I. Goldman1, A. Kreyssig, K. Prokěs, D.K. Pratt, D.N. Argyriou, J.W. Lynn, S. Nandi, S.A.J. Kimber, Y. Chen, Y.B. Lee, G. Samolyuk, J.B. Leão, S.J. Poulton, S.L. Bud'ko, N. Ni, P.C. Canfield, B.N. Harmon, and R.J. McQueeney,

arXiv:0811.2013v1.

[29] P.C. Canfield, I.R. Fisher, J. Cryst. Growth 225, 155 (2001).

[30] S. Muto, S. Takeda, R. Oshima, and F.E. Fujita J. Phys.: Condens. Matter 1, 9971 (1989).

[31] L.T. Romano, M.G. Smith, H. Oesterreicher, and R.D. Taylor, Phys. Rev. B 45 8042 (1992).

[32] Y. Zhu, M. Suenaga, and A.R. Moodenbaugh, Ultramicroscopy 37, 341 (1991).

[33] S. Semenovskaya and A. G. Khachaturyan, Phys. Rev. B 46, 6651 (1992).

Figure captions

Fig. 1. Temperature-dependent electrical resistivity of ThFe$_2$As$_2$ (Th=Sr, Ca). The data have been offset in the vertical direction for clarity.

Fig. 2. (a) Crystal structure of ThFe$_2$As$_2$ (Th=Sr, Ca, Ba). (b) Top view of the Fe layer and Th layer.

Fig. 3. High-resolution TEM image taken along the [100] zone-axis direction at room temperature, displaying the atomic structural features of SrFe$_2$As$_2$. The structural model is superimposed on the image, and the black rectangle indicates a unit cell. The inset shows a simulated TEM image.

Fig. 4. Bight field images (a, c) and corresponding electron diffraction patterns (b, d) taken along the [001] zone-axis direction at 300K and 100K, respectively. The black circle indicates the selected area for electron diffraction.

Fig. 5. (a) Typical bright-field TEM image illustrating the twining lamella in a SrFe$_2$As$_2$ crystal. (b) Schematic structural model illustrating twins within a Fe layer. TB denotes twin boundary. a´ (b´), parallel to the [110] ([-110]) direction in tetragonal phase, corresponds to the lattice parameter in the orthorhombic phase. (c) The [001] zone-axis electron diffraction pattern obtained from the area shown in (a). The inset shows enlarged (400) Bragg spot with unambiguous spot splitting along the <110>$_{orth}$ direction. (d) Schematic illustration of the electron diffraction pattern.

Fig. 6. Bright field images of CaFe$_2$As$_2$ taken at room temperature, showing a pseudo-periodic modulation (a) and a tweed structure (b). The insets show the fast Fourier transformation (FFT) for the TEM images.

Fig. 7. Electron diffraction patterns of CaFe$_2$As$_2$ taken along the [001] zone axis at 300K (a) and 100K (b), respectively. The inset shows enlarged (600) Bragg spot with unambiguous spot splitting along the <110>$_{orth}$ direction.

Fig. 8. (a) Bright field image of CaFe$_2$As$_2$ taken at 100K. (b) Schematic illustration of a mixture of the tweed and twin domain. The white and gray bands are the twin domains due to the tetragonal-orthorhombic phase transition.

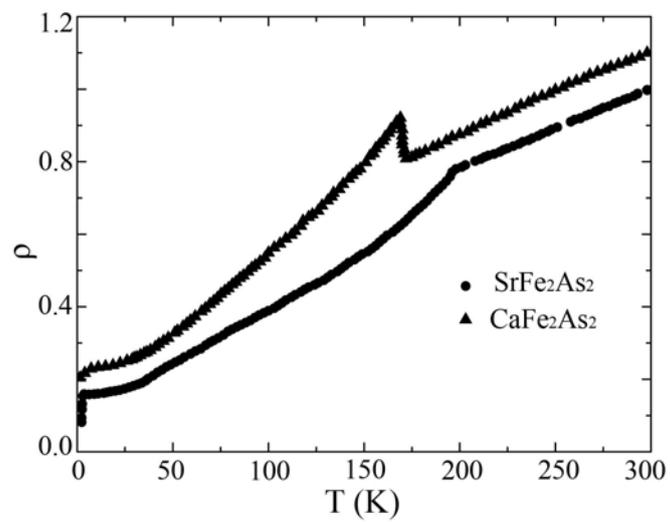

Fig. 1.

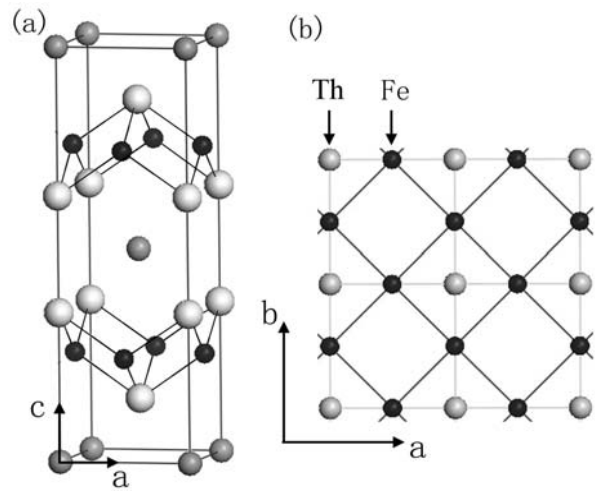

Fig. 2.

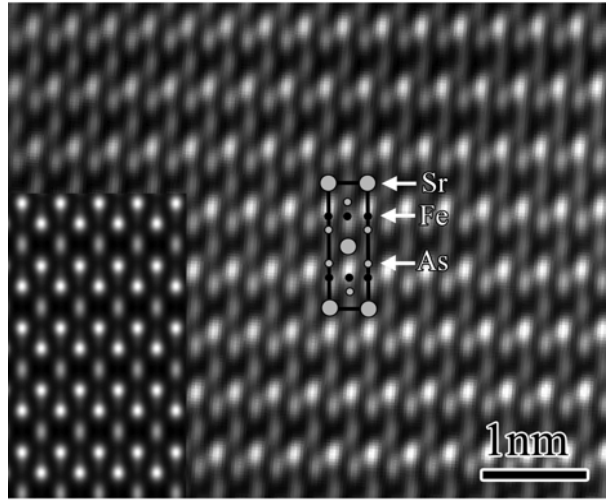

Fig. 3.

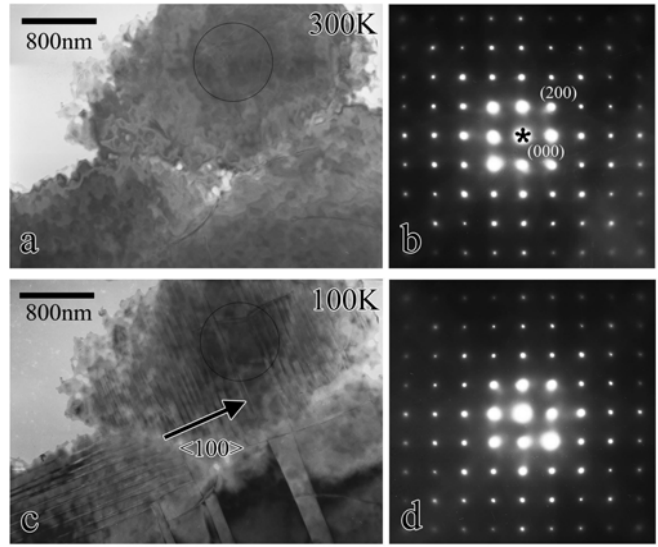

Fig. 4.

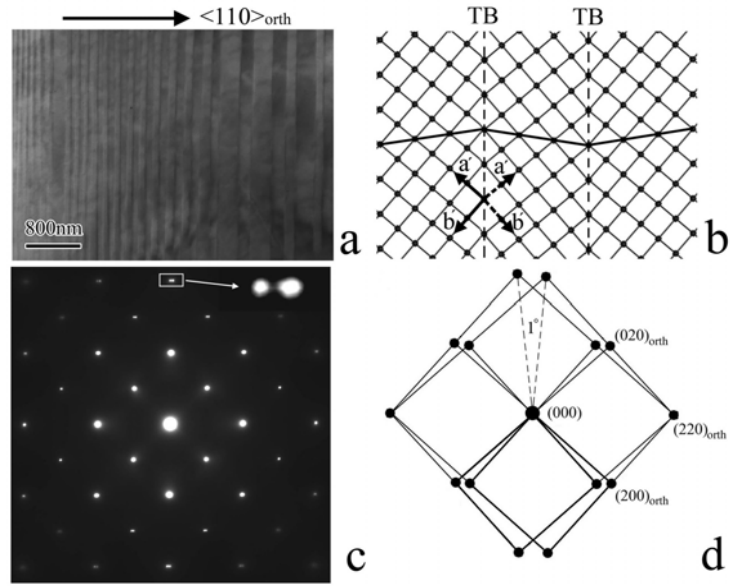

Fig. 5.

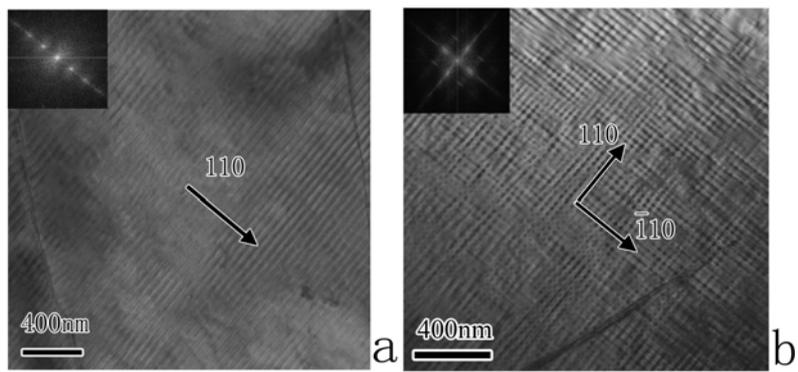

Fig. 6.

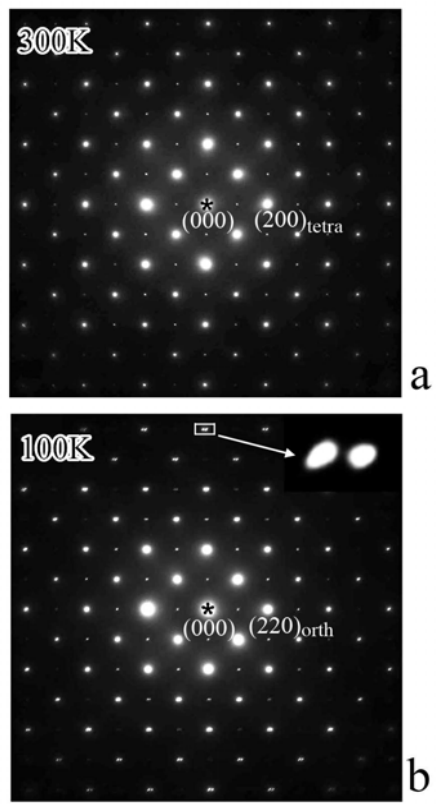

Fig. 7.

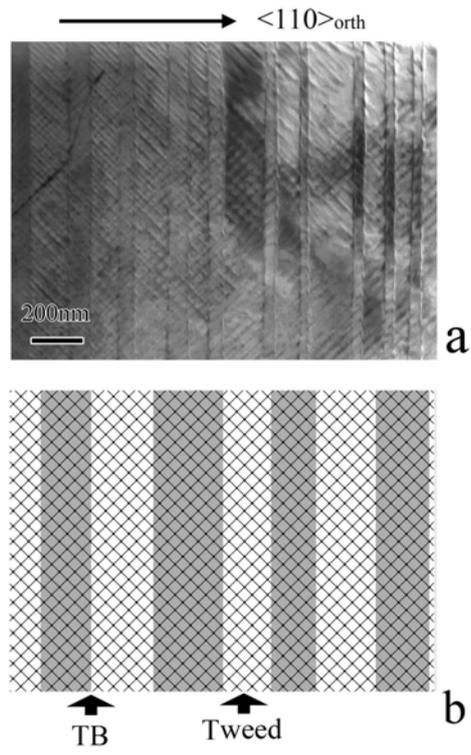

Fig. 8.